\documentclass[aps,twocolumn, pre, 
longbibliography, notitlepage]{revtex4-2}
\usepackage{bm,amsmath,amssymb,graphicx, mathtools,multirow,gensymb}
\usepackage{svg}
\usepackage[colorlinks=true,linkcolor=MidnightBlue,urlcolor=black,
citecolor=MidnightBlue,anchorcolor=MidnightBlue]{hyperref}
\usepackage[dvipsnames]{xcolor}
\usepackage{comment}
\begin{document}
%%%%%====================================
%%%%%====================================
%%%%%====================================
\title{Emergent flocking dynamics in chemorepulsive active colloids: \\Interplay of disorder and noise}
%%%%%====================================
%%%%%====================================
\author{Sagarika Adhikary} 
\thanks{a.sagarika@physics.iitm.ac.in} 
\affiliation{Department of Physics, Indian Institute of Technology Madras, Chennai 600036, India}
\author{Rajesh Singh} 
\thanks{rsingh@physics.iitm.ac.in} 
\affiliation{Department of Physics, Indian Institute of Technology Madras, Chennai 600036, India}
%%%%%====================================
%%%%%====================================
\begin{abstract}
Recent studies of active colloidal matter have revealed that a global polar order can arise from chemorepulsive interactions among particles without any explicit alignment interaction between them. In this work, we investigate such chemically interacting active colloids in the presence of quenched disorder, where a fraction of particles are randomly pinned in space. These pinned particles are restricted to rotational motion while remaining chemically coupled to the mobile population. In addition, angular noise is incorporated into the rotational dynamics to capture stochastic effects. To elucidate the interplay of quenched disorder and noise, we construct phase diagrams based on polar order and its fluctuations, and systematically analyze the associated disorder- and noise-driven phase transitions. Surprisingly, we find that the phase transition driven by the noise is significantly dependent on the density of the particles, whereas such a density-dependence is not present when the control parameter is the pinning fraction. The finite-size effects on these transitions are also examined. We demonstrate that an effective interaction range, governed by the coefficient related to screening of the chemorepulsive interaction, plays a crucial role in collective behavior. When the effective interaction range is much smaller than the system size, the system exhibits density band formation, a feature absent in the long-range interaction regime. Moreover, near the transition point, the order parameter distribution becomes bimodal for the case of short-range interaction. 
\end{abstract}
%%%%%====================================
%%%%%====================================
\maketitle
%%%%%====================================
%%%%%====================================
\section{Introduction}
%%%%%====================================
Active matter systems are collections of many active particles. An active particle dissipates energy from its
environment to create mechanical disturbance around it and/or to self-propel. 
Active matter systems \cite{ramaswamy2010mechanics, marchetti2013hydrodynamics,vrugt2025exactly} demonstrate a rich phenomenology due to
various interactions, disorder, and stochastic perturbations. 
Many studies have explored the collective dynamics of active matter, examining different types of inter-particle
interactions, including mechanisms of velocity alignment \cite{vicsek1995novel, solon2015pattern, chate2008collective, chen2024emergent, martin2018collective, sese2018velocity}, non-reciprocal interaction \cite{martin2025transition,gupta2022active,loos2023long,saha2025effervescence,kreienkamp2024dynamical,barberis2019phase}, history-dependent interactions \cite{kumar2023emergent, subramaniam2024rigid}, chemo-tactic interactions \cite{pohlStarkPRL2014, subramaniam2025minimal}, and others \cite{liebchen2019interactions,hiraiwa2020dynamic, grossmann2013self}. Although these advances have been made, the interplay between different interaction mechanisms and disorder in shaping collective motion across complex active environments remains an active area of research \cite{bechinger2016active,gompper20252025}. Various studies have examined collective behavior in active systems at increasing levels of complexity, including different forms of disorder \cite{morin2017distortion, stoop2018clogging, chepizhko2013diffusion, peruani2018cold, sandor2017dynamic, olsen2021active, siebers2023exploiting}, quenched disorder \cite{duan2021breakdown, das2018polar}, the presence of obstacles or external perturbations \cite{codina2022small, mokhtari2017collective, adhikary2021effect}, and population heterogeneity \cite{adhikary2022pattern, rouzaire2025activity, tang2025reentrant, jhajhria2025kinetics}. However, considerably less is known about the influence of complexity in chemically interacting active colloids, where particles interact through self-generated chemical fields rather than explicit velocity alignment. In particular, how quenched disorder modifies the onset and robustness of flocking \cite{toner2024physics,vicsek2012collective} in chemically interacting active systems remains to be systematically investigated. Conversely, active systems can be influenced by a variety of perturbations which are stochastic in nature. One notable example is angular, or rotational noise, which is often examined in studies of flocking transitions in self-propelled particle systems \cite{chepizhko2013optimal, shaebani2020computational}. 
The rich phenomenology emanating from the interplay between quenched disorder and the stochastic nature of interactions is an exciting area of current research.
%%%%%%%%%%%%%%%%%%%%%%%%%%%%%%%%%%%%%%%%%%%%%%%%%%%%%%%%%%%%%%%%%%%%%%%%%%%%%%
  \begin{figure*}[t]
     \centering
     \includegraphics[width=0.9\textwidth]{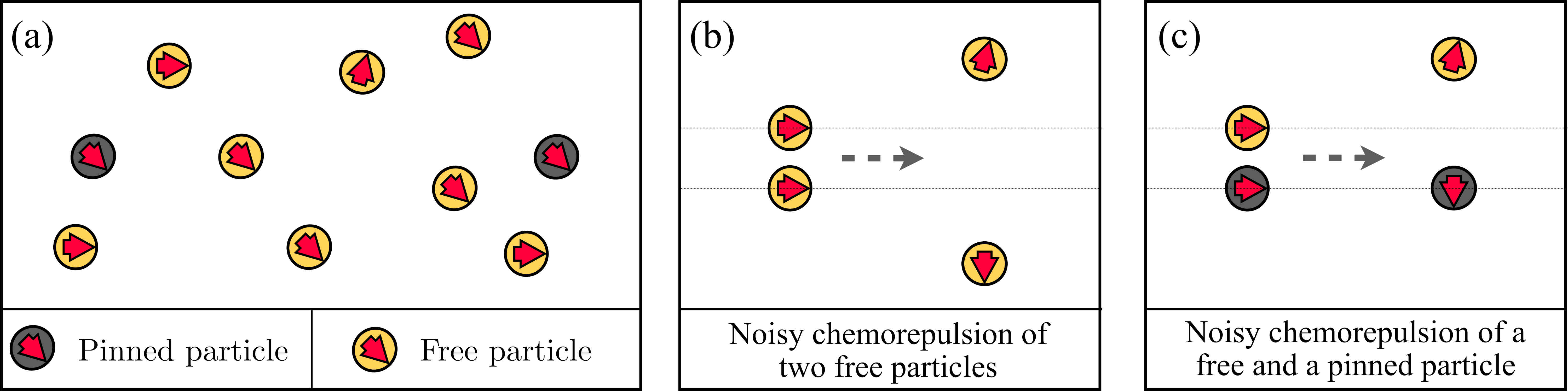}
     \caption{\label{phase_D1}
Schematic description of the model is shown in panel (a). Free and pinned active particles are randomly distributed initially in a two-dimensional space. Free and pinned particles are represented, respectively, by yellow and gray colour, red arrows indicates their orientation. 
(b) Noisy chemorepulsive 
interaction between two free active particles.
(c) Noisy chemorepulsive interaction between a 
free active particle 
and  a pinned active particle.}
 \label{schematic}
 \end{figure*}
%%%%%%%%%%%%%%%%%%%%%%%%%%%%%%%%%%%%%%%%%%%%%%%%%%%%%%%%%%%%%%%%%%%%%%%%%%%%%%

The question of whether polar order (flocking) can emerge intrinsically from particle-based dynamics, 
without any prescribed alignment interaction, has become a topic of growing interest 
\cite{das2024flocking, baconnier2025self, caprini2023flocking, subramaniam2025minimal, kumar2023emergent}. 
Studies of phoretic motion in active particles, driven by self-generated chemical fields, have shown that attractive torques can lead to the collapse of particles into dense clusters \cite{pohlStarkPRL2014}. However, comprehensive investigations of repulsive torques in such systems remain limited. Recently, a model incorporating long-range chemorepulsive interactions demonstrated that a flocking phase can emerge even in the absence of explicit alignment interactions between particles \cite{subramaniam2025minimal}. In this case, collective motion arises solely from the combined effects of repulsion of the excluded-volume at short-range and repulsive torques at long-range. Most studies of self-propelled particles are conducted in homogeneous environments composed of monodisperse constituents. In contrast, real active systems typically exhibit significant complexity and heterogeneity. A related study on pinning disorder has revealed a novel flocking transition, governed by a control parameter related to the strength of the disorder \cite{adhikary2025flocking}. In the present work, we investigate chemically interacting active colloids, where a fraction of the colloids are randomly pinned in space, allowing them to rotate only while still interacting with other particles, without any short-range excluded volume repulsion. In these systems, chemical interactions are long-ranged, but the presence of screening could effectively render them short-ranged. In addition, in the current model, the angular noise is set significantly higher. We then explore how quenched disorder and noise together influence the flocking transition in such systems. This is due to the fact that disorder and noise are often expected to endow the system with nontrivial emergent properties \cite{chepizhko2013optimal,duan2021breakdown, bechinger2016active}. In this work, we systematically investigate spatial order, flocking behavior, phase transitions, and finite-size effects. We further characterize the system through analyses of number-density fluctuations and pair correlation functions for both effectively short- and long-range interactions in the presence of quenched disorder and angular noise.
 
The remainder of the paper is structured as follows. In section \ref{sec:Model}, we describe our model of chemorepulsive active particles in the presence of pinning disorder and angular noise. In this section, the dynamics of the chemical field, the equation of motion, the derivation of dimensionless parameters, and the simulation details are described, respectively. In section \ref{sec:results}, We present results from particle-based numerical simulations by constructing phase diagrams in the absence of noise and disorder, and by characterizing the properties of the resulting phases. Then, an extensive study of flocking transition is presented for the case with long-ranged interactions present in the system, where various collective properties of the system and the robustness of the polar order are discussed. In Section \ref{sec:iv}, we examine the polar order and phase transition in the regime of effectively short-range interactions and compare the results with those obtained for long-range interactions using order parameter distributions, finite-size scaling, and number-density fluctuations. Finally, Section \ref{sec:conc} summarizes the main findings, discusses them in the context of existing literature, and highlights possible directions for future research.

\section{Model}\label{sec:Model}
We consider a system of $N$ chemically interacting active colloids of radius $b$, consisting of $N_m$ mobile particles and $N_p$ pinned particles such that $N_m + N_p = N$. The mobile particles $N_m$ are free to translate and rotate, while the pinned particles $N_p$ are fixed in position and can only rotate.
A schematic diagram of the model is shown in Fig.\ref{schematic}. A scenario of repulsive interaction between two free particles is shown in Fig.\ref{schematic}(b), when they are rotating away from each other and there is some asymmetry due to angular noise. The red arrow inside the particle indicates its orientation. For a free (yellow) and pinned (gray) particle interaction, the pinned particle only rotates (no translation), which also includes noise, as shown in Fig.\ref{schematic}(b). In Fig.\ref{schematic}(c), the dynamics of free and pinned particles is shown.
We present the dynamical equations of the particles and the chemical field in the following.
%%%=============================================
\subsection{Dynamics of the particles}

We model the $i$th active particle as a colloidal particle located at $\mathbf r_i=(x_i,y_i)$, confined to move in two-dimensional continuous space. Each particle self-propels with a constant speed $v_s$,  along the directions $\mathbf e_i =(\cos\theta_i,\,\sin\theta_i)$, which is subject to noise. Here, $i=1,2,3,\dots,N$ and $\theta_i$ denote the angle that the orientation vector $\mathbf e_i$ makes with the positive $x$-axis.
The dynamics of the $i$th particle is governed by:
\begin{align}
\label{eq:mainLE}
     \dot{\mathbf r}_i  &=  v_s \mathbf e_i +\zeta_t\, \mathbf J_i,
     \qquad \qquad \,\,i=1,\dots N_m,\\ 
     \dot{ \mathbf e}_i&= \left[\zeta_r 
    \left(\mathbf e_i\times\mathbf {\mathbf{{J}}}_i 
    \right) + \bm \eta^r_i\right] \times \mathbf e_i    ,\,\, \qquad \forall i.\nonumber
\end{align}
Here $v_s$ is the self-propulsion speed of an isolated moving particle. On the other hand, the pinned particles can not move such that:    
$\dot {\mathbf r}_{i} = 0$
for
$i=N_m+1,\dots , N_m+N_p$.
In the above, $\bm \eta^r_i$ is a white noise with zero mean and $2D_n$ variance with no temporal correlation.
As shown in Section~\ref{sec:concEv}, the phoretic flux $\mathbf{J}_i(t)$, arising from chemical interactions, is expressed in terms of the scalar field $c(\mathbf{r}, t)$. Using the expression for the flux, it follows that the term $\zeta_r$ in Eq.(\ref{eq:mainLE}) causes particles to rotate away from each other when $\zeta_r>0$ due to chemical interactions between particles. Similarly, the term proportional to $\zeta_t$ induces repulsion in the positional dynamics when $\zeta_t>0$. In this study, both $\zeta_r$ and $\zeta_t$ are positive, and therefore the system is referred to as \textit{chemorepulsive}. 
%%%=============================================
%%%=============================================
%%%=============================================
\subsection{Dynamics of the chemical field}\label{sec:concEv}
%%%=============================================
%%%=============================================
In the presence of a chemical field, the phoretic (chemical) flux $\mathbf{J}_i(t) 
$ is given in terms of the chemical field $c(\mathbf r, t)$, which represents the concentration of chemicals (for example, filled micelles in an oil-emulsion system \cite{hokmabad2022chemotactic, kumar2023emergent}).
The chemical field is obtained by modeling each particle as a source of the chemical field with an emission rate $\lambda_0$, while the field undergoes a uniform decay at a rate $\lambda_d$.
Consequently, the phoretic field $c(\mathbf r, t)$ evolves in time as
\begin{align}
     D_{c}\nabla^2 c(\mathbf r, t) + \sum_{i=1}^N\lambda_0\, \delta(\mathbf r- \mathbf r_i)
-\lambda_d c= 0.
\end{align}
 Here, $D_{c}$ is the diffusion coefficient. We note that our analysis assumes instantaneous chemical interactions between the colloidal surfaces \cite{saha2014clusters, soto2015self,ruckner2007chemically,singh2019competing}, corresponding to the steady-state limit of the diffusion equation for the chemical field. 
The expressions of the phoretic field $c(\mathbf r)$ is given in the following: 
\begin{align}
c(\mathbf{r}, t) &= \frac{\lambda_0}{4\pi D_c }\sum^N_{{j=1
    }} 
 \frac{ \exp\left[-\lambda\left(\mathbf{r} - \mathbf{r}_{j}\right) \right]
}
 {|\mathbf{r} - \mathbf{r}_{j}|}.
 \label{eq:curr_r2}
\end{align}
Here, $\lambda=\sqrt{\lambda_d/D_c}$ act as an inverse of screening length. If the value of $\lambda$ is small, the interaction range becomes long-range and if it is higher, an effective short-range interaction occurs. In this work, these two scenarios are investigated for the collective dynamics of chemically interacting active particles. We note that in this model the chemical field diffuses in an infinite three-dimensional half-space bounded by a planar surface. However, the particles themselves are confined to two-dimensional motion, moving within the plane of this bounding surface. Simulation details are given in appendix \ref{app:sim}.

\subsection{Dimensionless parameters}
The key dimensionless parameters are:
\begin{align}
    \Lambda_{r} &=\frac{\tau}{\tau_r}= \frac{\zeta_r }{b^3v_s}, \qquad
    \Lambda_{t} = \frac{\zeta_t}{b^4 v_s},\\
        \Lambda_n &=\frac{\tau}{\tau_n}
    =\frac{b \,D_n}{v_s},\qquad     n_p = \frac{N_p}{N_m+N_p} =  \frac{N_p}{N} .
\label{eq:lscales}
\end{align} 
Here, $b$ is the radius of a colloid, $\tau = {b}/{v_s}$ denotes the propulsion time scale of the moving particles, while
$\tau_r=b^4/\zeta_r$, characterizes the time scale associated with the deterministic rotational dynamics arising from phoretic interactions.  

$\Lambda_n$ measures the strength of the angular noise term, while $\tau_n=1/D_n$ is the time-scale of angular reorientation from noise. 
The strength of the disorder can be expressed in terms of the pinning fraction $n_p$, which is defined above.
We have an addition dimensionless parameter, the area fraction $\phi$. The area fraction $\phi$ is define as:
$\phi= {\left(N\pi b^2\right)}/{L^2}$. Here, $L$ is the size of the simulation box and $N$ is the total number of particles.

 %%%%%%%%%%%%%%%%%%%%%%%%%%%%%%%%%%%%%%%%%%%%%%%%%%%%%%%%%%%%%%%%%%%%%%%
  \begin{figure}[b]
     \centering
     \includegraphics[width=0.498\textwidth]{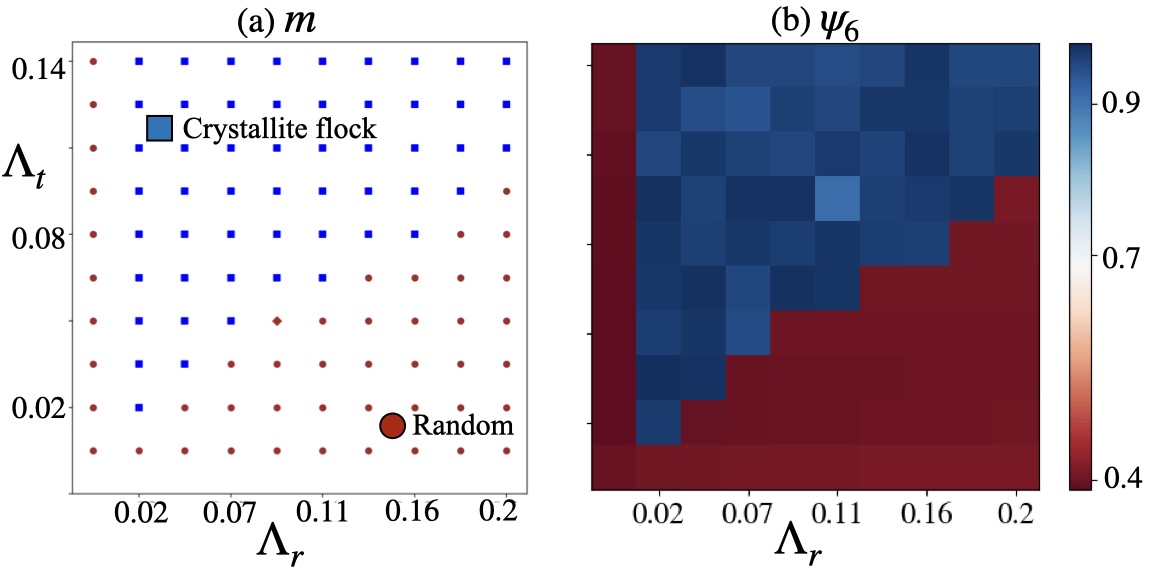}
     \caption{\label{phase_D1}
Without quenched disorder and noise: (a) Phase diagram of polar order; crystallite flock: $m>0.6$, Random: $m<0.1$ in the $(\Lambda_r,\Lambda_t)$ plane; (b) phase diagram of the sixfold bond-order parameter $\psi_6$ in the same plane. Fixed parameters: $L=200$, $\phi=0.1$, $n_p=0$, $\Lambda_n=0$, and $\lambda=0.01$.}
 \label{fig02}
 \end{figure}
%%%%%%%%%%%%%%%%%%%%%%%%%%%%%%%%%%%%%%%%%%%%%%%%%%%%%%%%%%%%%%%%%%%%%%%%%
%%%%%%%%%%%%%%%%%%%%%%%%%%%%%%%%%%%%%%%%%%%%%%%%%%%%%%%%%%%%%%%%%%%%%%%%%%%%%%%%%%%
 \begin{figure*}[t]
     \centering
     \includegraphics[width=0.99\textwidth]{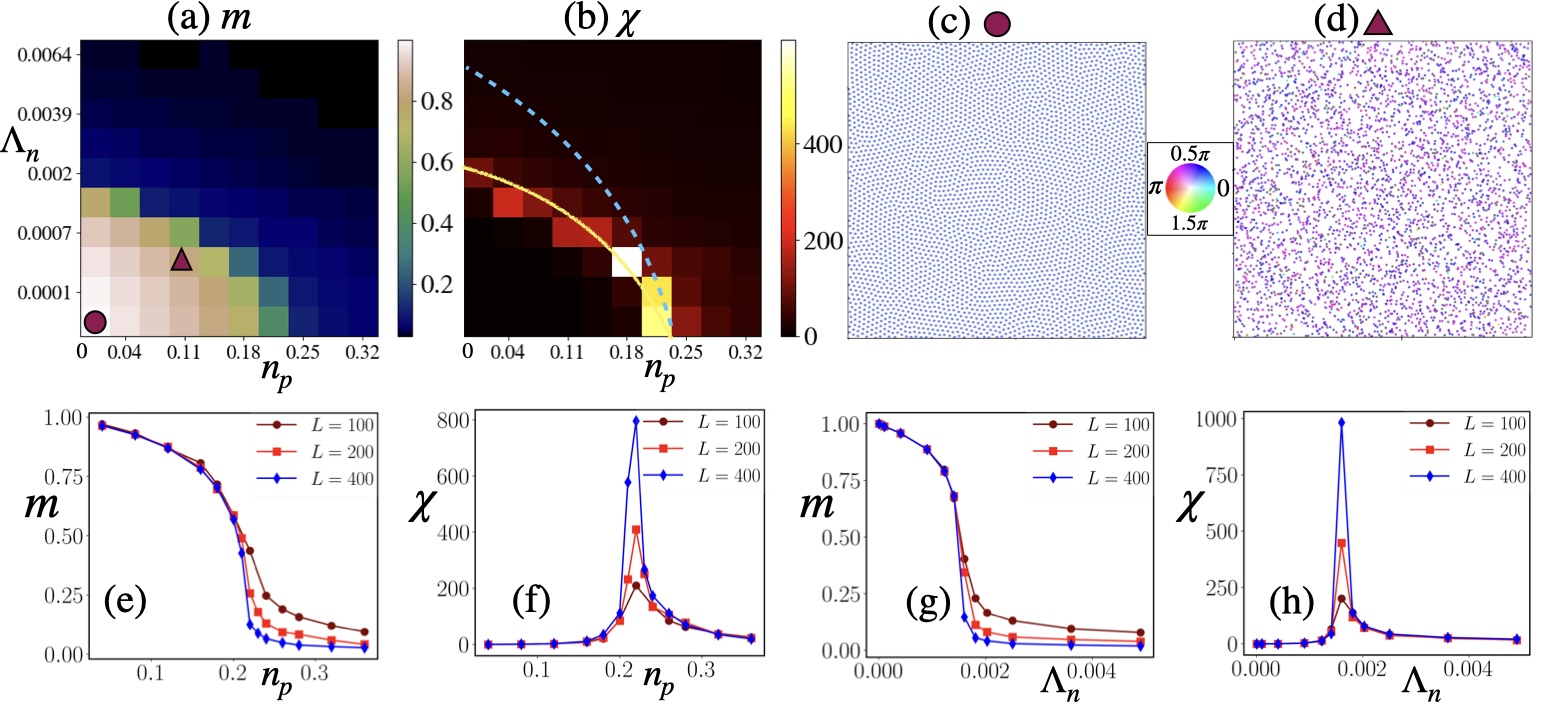}
     \caption{
Here, $\lambda=0.01$. (a) Phase diagram of $m$ in the $(n_p,\Lambda_n)$ plane, corresponding (b) fluctuations of order parameter ($\chi$). Fixed parameters: $L=200$, $\Lambda_r=0.1$, $\Lambda_t=0.1$, $\phi=0.1$. The continuous orange colored line in (b) shows the transition line approximately. The blue dotted line represents transition line for higher density with $\phi=0.36$. Steady-state configuration with (c) $\Lambda_n=0,n_p=0$, (d) $\Lambda_n=0.0003,n_p=0.1$, for total number of particles $N=4000$. Pinned particles are marked in gray color and direction of motion of the free particles follows the color scheme as shown in the middle of (c) and (d). Finite size effects on phase transition (with $\Lambda_n=0$): (e) order parameter ($m$), and (f) fluctuation ($\chi$) versus disorder ($n_p$). Finite size effects on phase transition (with $n_p=0$): (g) order parameter ($m$), and (h) ($\chi$) versus disorder ($\Lambda_n$). System sizes: $L=100,200$ and $400$, area fraction $\phi=0.1$.}
 \label{fig03}
 \end{figure*}
%%%%%%%%%%%%%%%%%%%%%%%%%%%%%%%%%%%%%%%%%%%%%%%%%%%%%%%%%%%%%%%%%%%%%%%%%%%%%%%%%%
\section{Results}\label{sec:results}
\subsection{Order parameters}
The order parameter characterizing the flocking transition is the global polarization $m$ of the particles \cite{vicsek1995novel,subramaniam2025minimal,adhikary2022pattern}, which is defined as:
\begin{align}
    m = \langle M \rangle_{ss},\qquad     M = \bigg| \frac1N \sum_{i=1}^N \mathbf{e}_i \bigg|
\end{align}
The instantaneous mean polarization $M$ at each time step is given by the average orientation of all particles. The angular brackets denote the averaging over the steady state of the dynamics. Details of statistical averaging are provided below in Appendix \ref{app:statAvg}. Global polarization $m$ serves as an order parameter that characterizes the flocking transition in terms of orientational order. A phase diagram is presented in Fig. \ref{fig02}(a) in the ($\Lambda_r,\Lambda_t$) plane, where we define $m>0.6$ as an ordered or flocking phase and $m<0.1$ as a disordered or random phase. In this case, no intermediate values of $m$ are found with the corresponding points in the phase plane.

Next, we study the local sixfold bond-order parameter $\psi_i$ for each particle and the related global bond-orientational order parameter $\psi_6$. They are determined following \cite{nelson1979, chaikin1995principles}:
\begin{align}
    \psi_6= \frac{1}{N} \sum_{i}^{N} \psi_{i},
    \qquad
    \psi_{i} =  \frac{1}{{N^n_i}} \sum_{j}^{{N^n_i}} e^{i6\theta_{ij}}.
\end{align}
Here, $\theta_{ij}$ denotes the angle between particles $i$ and particles $j$, and ${N^n_i}$ is the number of nearest neighbors of particles $i$ (the Voronoi tessellation method is used). 
The observable $\psi_6$ is determined by averaging over $1000$ distinct configurations in the steady state. The corresponding global six-fold bond order is shown in Fig. \ref{fig02}(b) in the ($\Lambda_r,\Lambda_t$) plane. When there is no quenched disorder or noise in the system, the polar ordered flocking phase is highly ordered in nature; this phase can be referred to as crystallite flock. In this phase, pair correlation shows distinctive peaks of crystalline order, which is shown in Fig.\ref{pair_corr}. In the present model, next we include quenched disorder and angular noise and investigate its interplay on emergent flocking dynamics. \\

\begin{figure*}[t]
     \centering
     \includegraphics[width=0.98\textwidth]{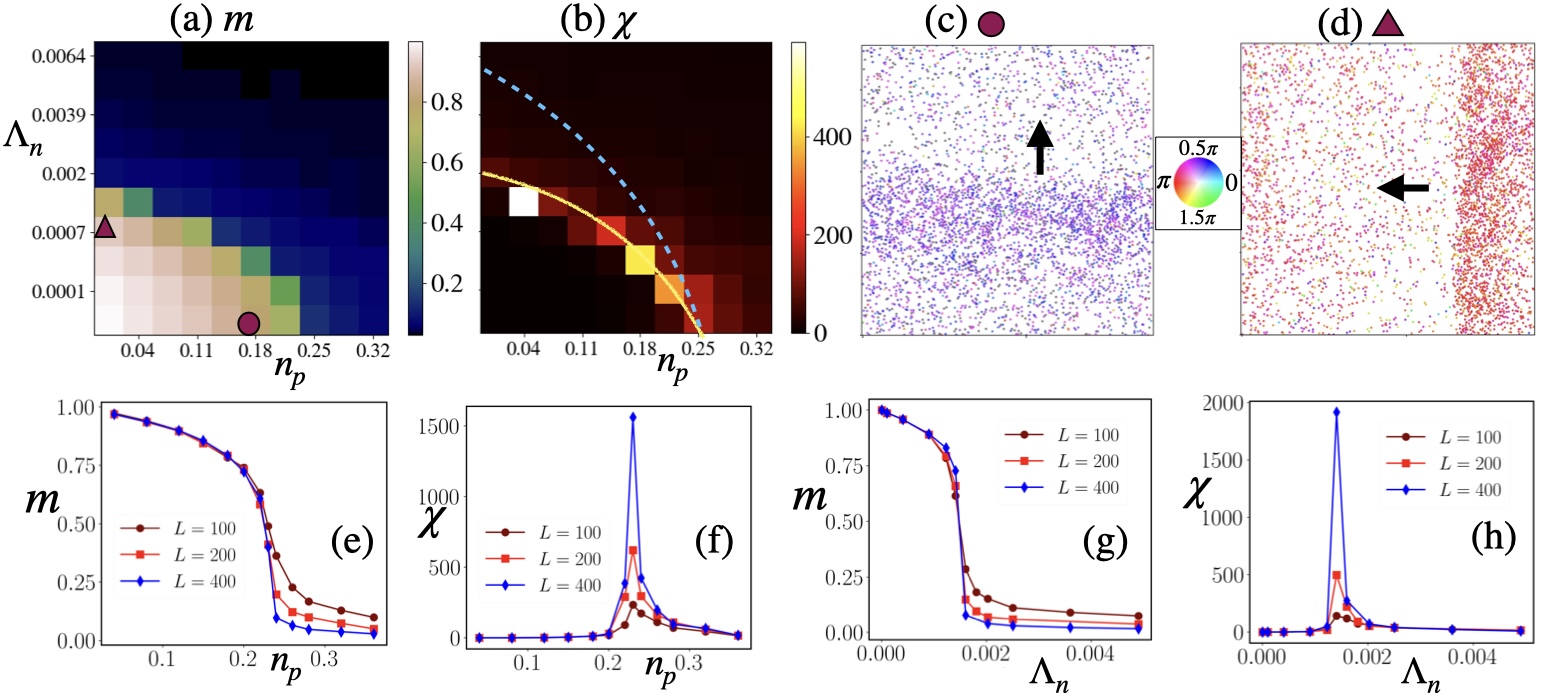}
     \caption{
Here $\lambda=0.5$.
(a) Phase diagram of $m$ in the $(n_p,\Lambda_n)$ plane, corresponding (b) fluctuations of order parameter ($\chi$). Fixed parameters: $L=200$, $\Lambda_r=0.1$, $\Lambda_t=0.1$, $\phi=0.1$. The continuous orange curve is fitted approximately to show the transition line. The blue dotted line in (b) represents transition line for higher density with $\phi=0.36$. Steady-state configuration with (c) $\Lambda_n=0.0008,n_p=0$, (d) $\Lambda_n=0,n_p=0.16$, for total number of particles $N=4000$. Pinned particles are marked in gray color and direction of motion of the free particles follows the color scheme as shown by the color bar. Finite size effects on phase transition (with $\Lambda_n=0$): (e) order parameter ($m$), and (f) fluctuation ($\chi$) versus disorder ($n_p$). Finite size effects on phase transition (with $n_p=0$): (g) order parameter ($m$), and (h) fluctuation $\chi$ versus disorder ($\Lambda_n$). System sizes: $L=100,200$ and $400$; area fraction $\phi=0.1$. }
\label{fig04}
\end{figure*}

\subsection{Phase transition: role of disorder and noise}\label{sec:ap1}
Now, keeping the parameters $\Lambda_r=\Lambda_t=0.1$ fixed, we vary the pinning fraction ($n_p$) and the noise ($\Lambda_n$). A corresponding phase diagram in the parameter plane ($n_p, \Lambda_n$) is presented in Fig.\ref{fig03}. The polar order parameter is shown in Fig.\ref{fig03}(a). The order is sustained only to some degree of disorder ($n_p$). The flocking phase is observed up to a noise strength ($\Lambda_n \approx 0.0016$). 
By analogy with equilibrium systems, we further determine the order parameter fluctuation $\chi$, which can be estimated as
\begin{equation}
  \label{c2d5}
  \chi = L^2 \left[\langle M^2\rangle - \langle M\rangle^2\right]
\end{equation}
The fluctuation $\chi$ is presented in Fig.\ref{fig03}(b). Near the flocking transition, $\chi$ becomes maximum due to the high fluctuation in the order parameter. From the maximum values in $\chi$, the phase transition line can be easily observed and marked with a continuous orange curve. These results are obtained for a fixed area fraction $\phi=0.1$. Similarly, for a higher density with an area fraction $\phi=0.36$, the transition line is also measured and shown in Fig.\ref{fig03}(b) with a blue-dotted curve. An interesting observation is that the phase transition with the noise parameter $\Lambda_n$ is significantly dependent on the density of the particles. However, such a dependence is not present when the control parameter is the pinning fraction $n_p$. Furthermore, the phase points shown in Fig.\ref{fig03}(a) are presented in (c)-(d) in terms of the morphology of the system in steady-state. The particle orientations are color-coded according to their angle with respect to the positive x-direction, as shown in the middle of Fig.\ref{fig03}(c) and (d). When the system is free of both disorder and noise, a crystallite flock forms (Fig.\ref{fig03}(c)). A small amount of pinning disorder and angular noise transform the spatial configuration into a liquid flock. The resulting morphology in the ordered state, near the flocking transition, is shown in Fig.\ref{fig03}(d). Small flocks of particles are found to travel in space, and the density band is not observed in the system. It should be noted that similar morphology is also observed near the transition, when only noise $\Lambda_n$ or disorder $n_p$ is varied.\\

Furthermore, we determine the finite size effect of the flocking transition, varying the control parameters $n_p$ and $\Lambda_n$ separately. First, without any noise $\Lambda_n=0$, the order parameter $m$ is plotted with varying pinning fraction $n_p$. The three different size systems $L=100$, $200$, and $400$ are taken for this with a fixed area fraction $\phi=0.1$. As shown in Fig.\ref{fig03}(e), at the transition point ($\approx 0.22$), the system undergoes a polar ordered phase to a random phase. In Fig.\ref{fig03}(f), the fluctuation in order parameter $\chi$ is plotted with $n_p$, where they peaked at $n_p=0.22$. Similarly for a system with $n_p=0$, the order parameter $m$ is plotted with different $\Lambda_n$. Three sizes $L=100$, $200$, and $400$ systems are taken for this with a fixed area fraction $\phi=0.1$. As shown in Fig.\ref{fig03}(g) and (h), at the transition point ($\approx 0.0016$), the system undergoes a polar ordered phase to a random phase. Next, we will study the emergent collective motion with an effective short ranged chemorepulsive interaction with the co-efficient value $\lambda=0.5$ in the following section.

%%%%%%%%%%%%%%%%%%%%%%%%%%%%%%%%%%%%%%%%%%%%%%%%%%%%%%
%%%%%%%%%%%%%%%%%%%%%%%%%%%%%%%%%%%%%%%%%%%%%%%%%%%%%%%%%%
%%%%%%%%%%%%%%%%%%%%%%%%%%%%%%%%%%%%%%%%%%%%%%%%%%%%%%%%%%
\subsection{Impact of effective short-ranged interaction}
\label{sec:iv}
%%%%%%%%%%%%%%%%%%%%%%%%%%%%%%%%%%%%%%%%%%%%%%%%%%%%%%%%%%
%%%%%%%%%%%%%%%%%%%%%%%%%%%%%%%%%%%%%%%%%%%%%%%%%%%%%%%%%%
Long-range chemical interactions become effectively short if the screening is in a range much smaller than that of the system size. For the value of the coefficient $\lambda=0.5$, we have studied the collective dynamics keeping all other parameters the same as before. First, varying the pinning disorder $n_p$ and noise $\Lambda_n$, a phase diagram is determined in terms of polar order. It is shown in Fig.\ref{fig04}(a). The fluctuation in the order parameter is plotted in Fig.\ref{fig04}(b) and shows a maximum fluctuation near the transition (shown by the continuous orange curve). A transition line for higher density with $\phi=0.36$ is also plotted and shown with a blue-dotted curve, which shows a similar behavior as seen in the previous case with $\lambda=0.01$. A steady-state snapshot of the particle configurations near the phase transition (in the ordered phase) is shown in Fig.\ref{fig04}(c), where $\Lambda_n=0$ and $n_p=0.16$. The formation of a traveling density band-like structure can be seen in this case, where particles present in the band move in a direction (indicated by a black arrow) perpendicular to the band length. Then a steady-state configuration of another density band phase is shown in Fig.\ref{fig04}(d), where $n_p=0$ and $\Lambda_n=0.0008$. In this case, the traveling density band is more prominent, without any spatial disorder. In Vicsek-like models with local alignment rule, traveling density band formation occurs in the case of a discontinuous transition \cite{adhikary2021effect, vicsek2012collective, chate2008collective}. Hence, in the present model with effective short-range interaction ($\lambda=0.5$), the formation of the density band indicates a discontinuous transition with two-phase co-existence at the transition point.
%%
 %%%%%%%%%%%%%%%%%%%%%%%%%%%%%%%%%%%%%%%%%%%%%%%%%%% 
  \begin{figure}[t!]
     \centering
     \includegraphics[width=0.42\textwidth]{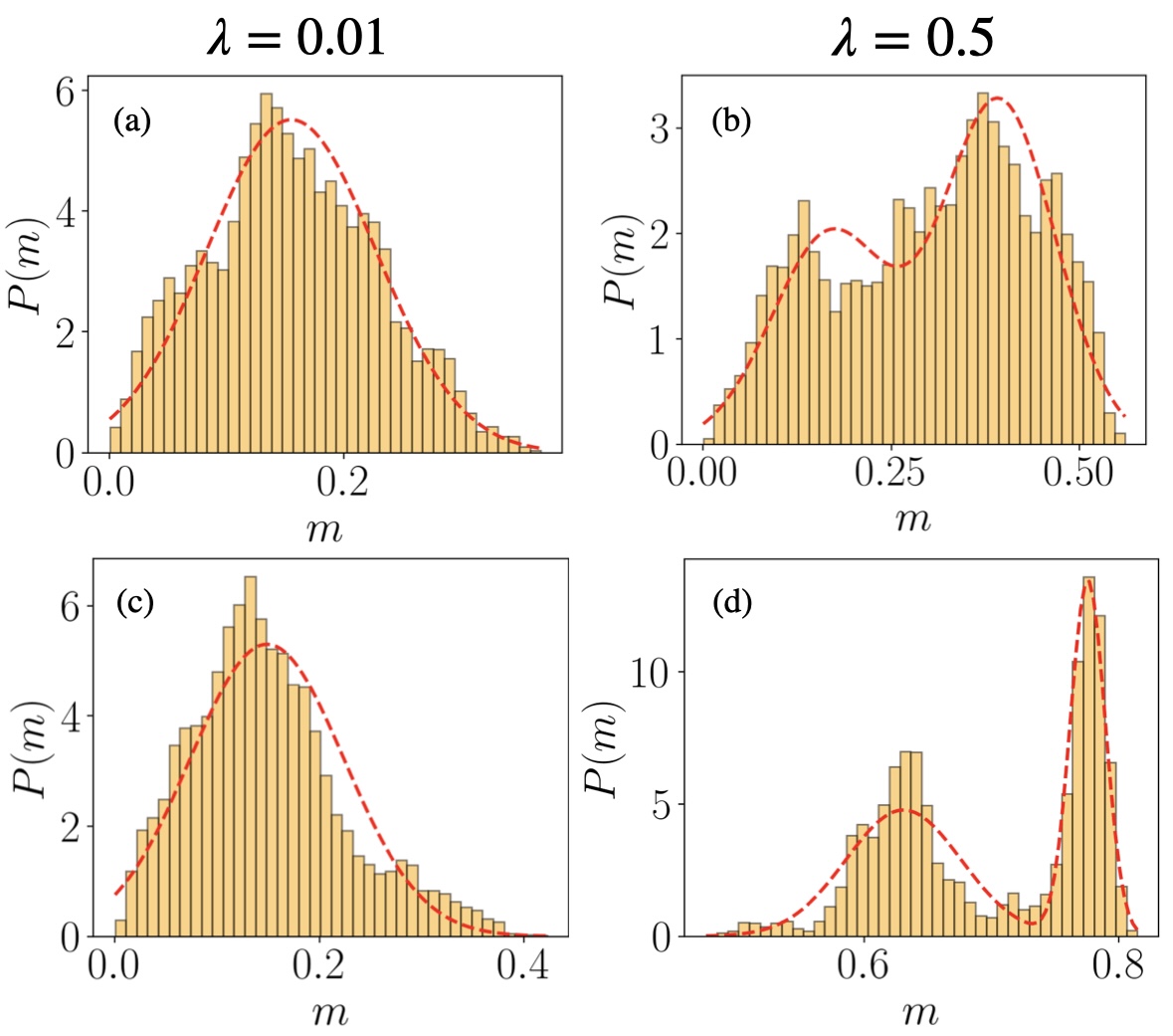}
     \caption{\label{fig05}
Probability Distribution of the order parameter $m$ at transition point.
Left panel: effective long-ranged interaction with co-efficient $\lambda=0.01$ (a) for control parameter $n_p(=0.22)$, (c) for control parameter $\Lambda_n(=0.0016)$. Right panel: effective short-ranged interaction with co-efficient $\lambda=0.5$ (b) for control parameter $n_p(=0.23)$, (d) for control parameter $\Lambda_n(=0.00144)$. System size $L=400$; area fraction $\phi=0.1$. }
 \end{figure}
%%%%%%%%%%%%%%%%%%%%%%%%%%%%%%%%%%%%%%%%%%%%%%%%%%%%%%

Furthermore, finite-size effects are also determined for the flocking transition with both control parameters $n_p$ and $\Lambda_n$. The phase transition plot in terms of the polar order parameter $m$ versus $n_p$ is shown in Fig.\ref{fig04}(e) for three different sizes of simulation boxes of length $L=100, 200$ and $400$. The corresponding order parameter fluctuations $\chi$ are shown in Fig.\ref{fig04}(f). From the peak of $\chi$, the transition point ($n_p=0.23$) can be determined. The transition point shifted to a slightly higher value of $n_p$ in this case than in $\lambda=0.01$. Similarly, the finite size effects of phase transition with the control parameter $\Lambda_n$ is also studied. The order parameter $m$ and corresponding fluctuation $\chi$ versus $\Lambda_n$ are shown in Fig.\ref{fig04}(g) and (h) for three different sizes of simulation boxes of length $L=100, 200$ and $400$. The $\chi$ plots peak at the transition point ($\Lambda_n=0.00144$), as shown in Fig.\ref{fig04}(h) The transition point shifted to a slightly lower value of $\Lambda_n$ in this case with respect to $\lambda=0.01$. More detailed comparison in these two range of interactions are discussed in the upcoming sections.

\subsubsection{Probability distribution of the order parameter $m$ for long and short-range interaction}
The distribution of the order parameter for this flocking transition behaves differently near the transition for long- and short range interactions. First, the probability distribution of $m$ for $\lambda=0.01$ is plotted in the left panel of Fig. \ref{fig05}. For the control parameter $n_p$, the distribution $P(m)$ is shown in Fig. \ref{fig05}(a) at the transition point $n_p=0.22$ and fitted with a distribution function (dotted red curve). The single humped distribution indicates that there are no two-phase co-existing in the dynamics. Similarly, for the control parameter $\Lambda_n$, the distribution $P(m)$ is shown in Fig. \ref{fig05}(c) at the transition point $\Lambda_n=0.0016$ and fitted with a distribution function, which also shows similar properties. Then the distribution $P(m)$ for $\lambda=0.5$ is plotted in the right panel of Fig. \ref{fig05}. For the control parameter $n_p$, the distribution $P(m)$ is shown in Fig. \ref{fig05}(b) at the transition point $n_p=0.23$ and fitted with a distribution function. There is a signature of double hump distribution in this case. It is more prominent for the control parameter $\Lambda_n$, and the distribution $P(m)$ is shown in Fig. \ref{fig05}(d) at the transition point $\Lambda_n=0.00144$. The double peak distribution of the latter case indicates dynamical two-phase co-existence at the transition point. From these observations we can state that when the interaction is in a much smaller length than the system size, active particles tend to form density band structure and the corresponding first-order flocking transition. \\

  %%%%%%%%%%%%%%%%%%%%%%%%%%%%%%%%%%%%%%%%%%%%%%%%%%%%%%%%%%%%
  \begin{figure}[b]
     \centering
     \includegraphics[width=0.42\textwidth]{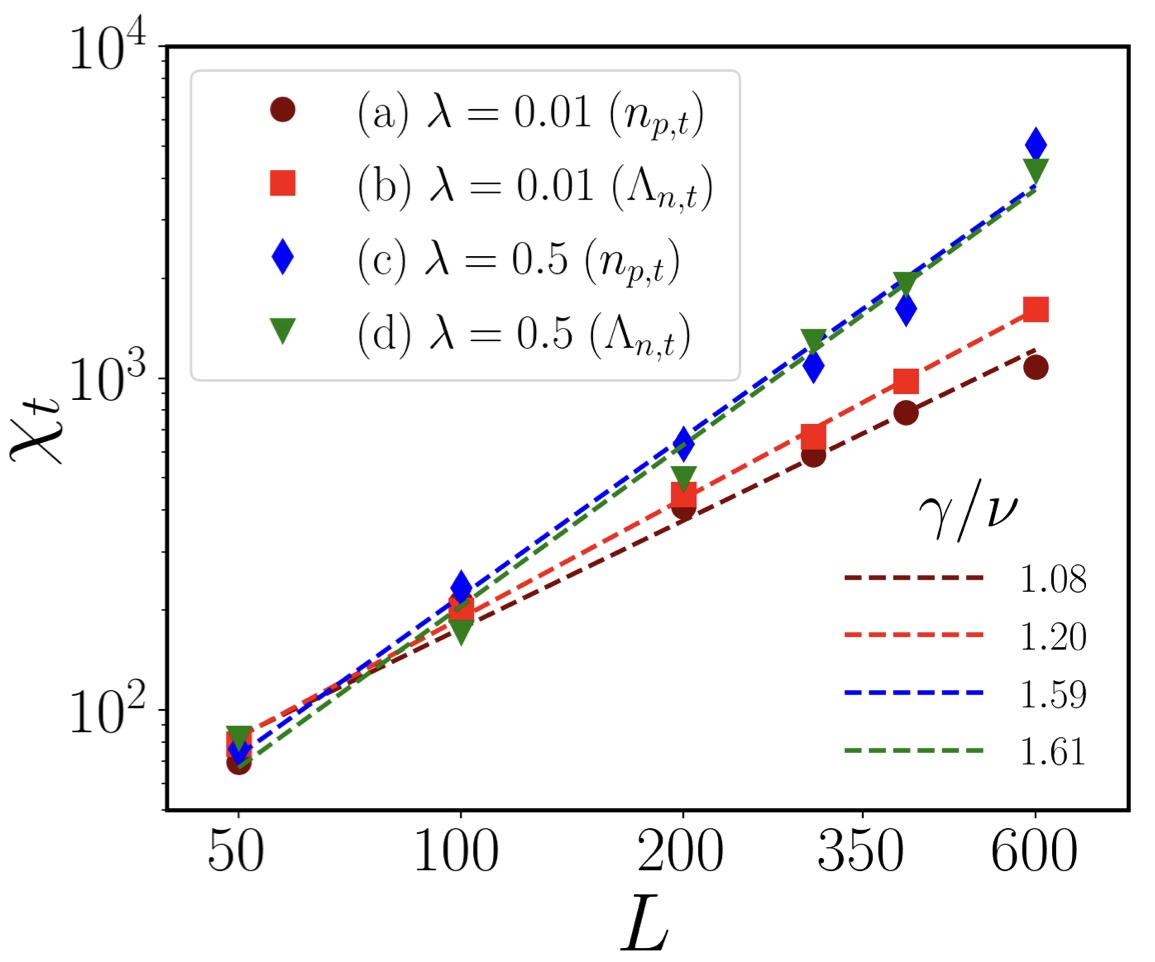}
     \caption{\label{scaling} Fluctuation of order parameter at the transition point ($\chi_t$) over different system of sizes $L=[50,100,200,300,400,600]$ for four different cases with different $\lambda$ and control parameters ($n_p$, and $\Lambda_n$).} 
 \end{figure}
 %%%%%%%%%%%%%%%%%%%%%%%%%%%%%%%%%%%%%%%%%%%%%%%%%%%%%%%%%%%%
  
\subsubsection{{Finite-size scaling}}
Following equilibrium critical phenomena
\cite{binder1987theory,christensen2005complexity}, a finite-size
scaling (FSS) relation for the order parameter can be assumed as
\begin{equation}
  \label{c2d7}
   m(p,L)=L^{-\beta/\nu}m_{0}[\epsilon L^{1/\nu}]
\end{equation} 
where $\epsilon=(p - p_{t})/p_{t}$ is the reduced control parameter ($n_p$ or $\Lambda_n$). Here, $p$ represents the control parameter $n_p$ or $\Lambda_n$ and the subscript $t$ represents the transition point. The exponent $\beta$ is the critical exponent of the order parameter, $\nu$ is the correlation length exponent, and $m_{0}$ is a scaling function. The order parameter distribution is defined as
\begin{equation}
  \label{c2pd}
  \mathbb{P}_L(M) = L^{\beta/\nu} \widetilde{\mathbb{P}}_L\left[M L^{\beta/\nu}\right]
\end{equation} 
where $\widetilde{\mathbb P}_L$ is a scaling function and $\langle M^n\rangle=\int M^n \mathbb P_L(M)dM$.
The FSS form of $\chi$ is then given by 
\begin{equation}
  \label{c2d8}
  \chi(p,L)= L^{\gamma/\nu}S_{0}[\epsilon L^{1/\nu}]
\end{equation}  
where $S_0$ is a scaling function and $\gamma/\nu=d-2\beta/\nu$ as
both $\langle M^2\rangle$ and $\langle M\rangle^2$ is 
$L^{-2\beta/\nu}$ and $d=2$. The exponent $\gamma$ is
defined as $\chi \sim \epsilon^{-\gamma}$. In the case of a first-order transition, the critical exponent $\beta$, related to the order parameter, goes to zero. As a consequence, $\chi$ should then scale as $\chi\sim L^d$, where $d$ is the dimension of space \cite{janke2003first,chate2008collective}. At the transition point, the finite-size scaling relation predicts $\chi_t \sim L^{\gamma/\nu}$. We plot the fluctuation measure $\chi_t$ as a function of the system size $L$ for four different cases in Fig.~\ref{scaling}, corresponding to the transitions shown in Figs.~\ref{fig03}(e)–(h) for $\lambda=0.01$ and Figs.~\ref{fig04}(e)–(h) for $\lambda=0.5$. For the effectively long-range interaction ($\lambda=0.01$), the estimated scaling exponent is $\gamma/\nu \approx 1.08$ when the control parameter is the pinning fraction $n_p$ (a), and $\gamma/\nu \approx 1.20$ when the control parameter is the angular noise strength $\Lambda_n$ (b). This difference suggests that quenched disorder and angular noise affect the flocking transition differently. For the effectively short-range interaction ($\lambda=0.5$), the corresponding effective exponents increase to $\gamma/\nu \approx 1.59$ for the disorder-driven transition (c) and $\gamma/\nu \approx 1.61$ for the noise-driven transition (d). The larger values of $\gamma/\nu$ indicate stronger fluctuations at the transition for the accessible system sizes, consistent with the tendency toward a sharper transition in the short-range interaction regime. \\

 %%%%%%%%%%%%%%%%%%%%%%%%%%%%%%%%%%%%%%%%%%%%%%%%%%% 
 %%%%%%%%%%%%%%%%%%%%%%%%%%%%%%%%%%%%%%%%%%%%%%%%%%% 
\subsubsection{Number fluctuations in particle density}
%%%%%%%%%%%%%%%%%%%%%%%%%%%%%%%%%%%%%%%%%%%%%%%%%%% 
 %%%%%%%%%%%%%%%%%%%%%%%%%%%%%%%%%%%%%%%%%%%%%%%%%%% 
  \begin{figure}[t]
     \centering
     \includegraphics[width=0.4\textwidth]{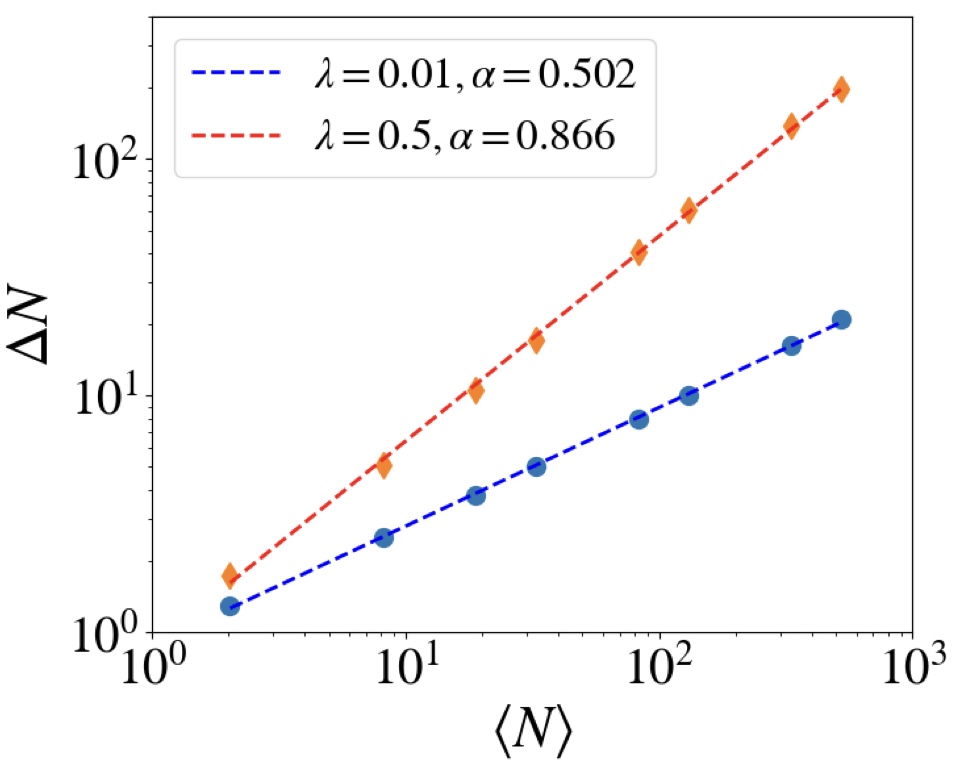}
     \caption{\label{gnf} Number fluctuations with two different effective long and short-ranged interaction with co-efficient $\lambda=0.01$ and $\lambda=0.5$ at the flocking phase with disorder $n_p(=0.1)$, and noise $\Lambda_n(=0.0001)$. System size $L=512$; area fraction $\phi=0.1$.} 
 \end{figure}

The ordered phase, in which active particles flock collectively along a preferred direction, is often associated with giant number fluctuations (GNF) \cite{ramaswamy2010mechanics,ginelli2016physics}. GNF refers to anomalously large density fluctuations, characterized by the scaling of the standard deviation of the particle number as $\Delta N \sim N^{\alpha}$, where the equilibrium expectation is $\alpha = 0.5$. We investigate these fluctuations for both effective long-range ($\lambda = 0.01$) and short-range ($\lambda = 0.5$) chemorepulsive interactions. For a fixed pinning fraction of $n_p = 0.1$ within the polar-ordered phase, the number fluctuations are shown in Fig.~\ref{gnf} for the system of size $L=512$. Here, a total number of $500$ steady-state configurations is taken to determine the number fluctuations. The measured exponents are $\alpha \approx 0.502$ for $\lambda = 0.01$ and $\alpha \approx 0.866$ for $\lambda = 0.5$. The long-range interaction regime exhibits fluctuations close to the equilibrium prediction, whereas the effectively short-range interaction regime displays pronounced giant number fluctuations. These enhanced density fluctuations are consistent with the emergence of traveling density bands and are characteristic of many flocking active matter systems \cite{ramaswamy2003active,narayan2007long, chate2006simple}.

\section{Summary and Conclusion}\label{sec:conc}
In this work, we investigate the interplay between pinning disorder and angular noise in the emergence of flocking behavior in chemorepulsive active particles. Particles interact via chemorepulsive forces and torques, which are long-ranged in nature and subject to a tunable screening length. We first characterize the phase behavior in the absence of disorder and noise by varying the strengths of the force and torque. In this regime, we identify two distinct phases: a crystalline flocking phase, exhibiting both global sixfold bond-orientational order and polar order, and a disordered phase in which both types of order are absent. We then fix the force and torque parameters within the crystallite flocking regime and systematically investigate how quenched disorder and angular noise influence the stability and dynamics of this ordered state.\\

We first analyze the system in the effectively long-range interaction regime by constructing phase diagrams based on the polar order parameter and its fluctuations. The corresponding flocking transition and finite-size effects are then analyzed. When angular noise is used as the control parameter, the transition is found to depend sensitively on the area fraction (particle density), with higher densities favoring the stability of the ordered flocking phase, consistent with the behavior observed in Vicsek-like systems with local alignment interaction. In contrast, when the pinning disorder is varied, the transition shows negligible dependence on density, indicating a fundamentally different role for the quenched disorder compared to noise. In the effectively long-range interaction regime, the ordered phase remains spatially homogeneous near the transition, and no traveling density bands are observed for the system sizes considered. Introducing strong screening reduces the interaction range relative to the system size and qualitatively changes the collective dynamics. Near the transition, the system develops traveling density bands, giving rise to dynamical phase coexistence. Correspondingly, the order-parameter distribution evolves from a single-peaked to a bimodal form, providing further evidence of dynamical phase coexistence near the transition. In addition, the short-range interaction regime exhibits enhanced number-density fluctuations, consistent with the emergence of giant number fluctuations commonly observed in active matter systems. Overall, our results highlight the distinct roles of the interaction range, quenched disorder, and angular noise in determining the collective behavior of chemically interacting active colloids.\\

In this paper, we have presented a minimal model to study the interplay of phoretic interactions, noise and disorder. At the same time, we have made several simplifying assumptions. First, we neglect hydrodynamic interactions, which are known to significantly influence the collective behavior and structure formation of active particles \cite{thutupalli2018flow, singh2016universal, zhou2026hydrodynamic, matas2014hydrodynamic, gidituri2026microswimmers}. Second, we assume a constant self-propulsion speed for all particles and neglect fluctuations in the chemical concentration field. Finally, the pinned particles are distributed randomly, without considering spatial correlations in the disorder. While these assumptions allow us to focus on the role of chemorepulsive interactions, quenched disorder, and stochastic noise on the emergent dynamics. Relaxing these assumptions would enable a more comprehensive understanding of the collective dynamics in chemically interacting active systems. In particular, incorporating hydrodynamic interactions between particles and studying the role of non-constant self-propulsion speeds  \cite{torrenegra2024self, martin2021statistical,caprini2022dynamics} represents a natural extension of the present model and an important avenue for future studies.

\section*{Supplementary Material}
The time evolution of the formation of the flocking pattern for different parameter values is attached as supplementary movies \cite{siText}. The free particles are colored according to their orientations (same as the one used in Fig.\ref{fig03} and Fig.\ref{fig04}). The pinned particles are marked as gray, irrespective of their orientations. The resulting dynamics of the collective pattern formation remain robust if we increase the size of the system.

\begin{itemize}
    \item \textbf{Movie I}.
Dynamics of the crystallite flock: Without any disorder and noise as shown in Fig. \ref{fig03}(c). Total number of particles $N=4000$, $\phi=0.1$, $\lambda=0.01$, $n_p=0$, and $\Lambda_n=0$.

\item \textbf{Movie II}.
Dynamics of the liquid flock: Formation of the liquid flock with disorder and noise as shown in Fig. \ref{fig03}(d). Total number of particles $N=4000$, $\phi=0.1$, $\lambda=0.01$, $n_p=0.1$, and $\Lambda_n=0.0003$.

\item \textbf{Movie III}.
Dynamics of the density band flock: Formation of the traveling density band only with disorder as shown in Fig. \ref{fig04}(c). Total number of particles $N=4000$, $\phi=0.1$, $\lambda=0.5$, $n_p=0.18$, and $\Lambda_n=0$.

\item \textbf{Movie IV}.
Dynamics of the density band flock: Formation of the traveling density band with noise as shown in Fig. \ref{fig04}(d). Total number of particles $N=4000$, $\phi=0.1$, $\lambda=0.5$, $n_p=0$, and $\Lambda_n=0.0008$.

\end{itemize}

\section*{Acknowledgments}  
We thank the two anonymous reviewers for comments, which led to improvement of our manuscript. SA acknowledges support from the National Postdoctoral Fellowship (SERB File number: PDF/2023/002096) provided by ANRF, Government of India.\\

\appendix

\section{Simulation details}
\label{app:sim}
The equations of motion, as described in \eqref{eq:mainLE}, are simulated using a Euler-Maruyama integrator. 
Following the definition of the phoretic field in Eq.\eqref{eq:curr_r2}, the expression of the phoretic flux $\mathbf J = -\left[\mathbf\nabla c(\mathbf r, t)\right]_{\mathbf r=\mathbf r_i}$ is given as
\begin{align}
     \mathbf{J}_i(t) &= \frac{\lambda_0}{4\pi D_c }\sum^N_{\substack{j=1\\ i\neq j}} 
 \frac{\mathbf  r_{ij} (\lambda r_{ij}+1) \exp (-\lambda r_{ij})}{
 r_{ij}^3 }.
\end{align}
The above is used in \eqref{eq:mainLE} to run simulations.
The time-stepping of the Euler–Maruyama integrator is taken as $dt=0.01$. Initially, both mobile and pinned particles are randomly distributed in the two-dimensional space, and their orientations are randomly assigned from a uniform distribution of angles [$-\pi,+\pi$] measured with respect to the positive $x$-axis. The pinned particles are then randomly selected and the resulting collective dynamics is analyzed. Periodic boundary conditions are applied along both the $x$-axis and the $y$-axis of a square box of length $L$. The area fraction $\phi$ is fixed at $\phi=0.1$ for most simulations, unless otherwise specified.  
A table of all the parameters used for the generated figures is presented in Table \ref{si_table_params}. 
\begin{table*}[t!]
	\centering
	\begin{tabular}[c]{|l|l|l|l|l|l|l|l|l|}
     	\hline
     	Figure no. & $\lambda$ & $N$ & $L$ & $\phi$ & $n_p$ & $\Lambda_t$ & $\Lambda_r$ & $\Lambda_n$ \\
     	\hline
     	2.(a)-(b) & $0.01$ & $1274 $ & $200$ &  $0.1$ & $0$  & $(0,0.14)$ & $(0,0.2)$ & $0$  \\
        \hline
        3.(a)-(b) & $0.01$ & $1274$ & $200$ & $0.1$ & $(0,0.32)$ & $0.1$ & $0.1$ & $(0,0.0064)$ \\
        \hline 
        3.(c) & $0.01$ & $4000$ & $354$ &  $0.1$ & $0$  & $0.1$ & $0.1$ & $0$  \\
        \hline 
        3.(d) & $0.01$ & $4000 $ & $354$ &  $0.1$ & $0.1$  & $0.1$ & $0.1$ & $0.0003$  \\ 
        \hline         
        3.(e)-(f) & $0.01$ & $(318-5096)$ & $(100-400)$ &  $0.1$ & $(0,0.36)$  & $0.1$ & $0.1$ & $ 0$  \\
        \hline
        3.(g)-(h) & $0.01$ & $(318-5096)$ & $(100-400)$ & $0.1$ & $0$ &$0.1$ & $0.1$  & $(0,0.005)$ \\
        \hline    
        4.(a)-(b) & $0.5$ & $1274$ & $200$ & $0.1$ & $(0,0.32)$ & $0.1$ & $0.1$ & $(0,0.0064)$  \\
        \hline    
        4.(c) & $0.5$ & $4000$ & $354$ &  $0.1$ & $0.18$  & $0.1$ & $0.1$ & $0$   \\
        \hline    
        4.(d) & $0.5$ & $4000$ & $354$ &  $0.1$ & $0$  & $0.1$ & $0.1$ & $0.0008$   \\
        \hline    
        4.(e)-(f) & $0.5$ & $(318-5096)$ & $(100-400)$ &  $0.1$ & $(0,0.36)$  & $0.1$ & $0.1$ & $ 0$   \\
        \hline            
        4.(g)-(h) & $0.5$ & $(318-5096)$ & $(100-400)$ & $0.1$ & $0$ &$0.1$ & $0.1$  & $(0,0.005)$   \\
        \hline          
        5.(a),(c) & $0.01$ & $5096$ & $400$ & $0.1$ & $0.22,0$ &$0.1$ & $0.1$  & $0,0.0016)$   \\
        \hline 
        5.(b),(d) & $0.5$ & $5096$ & $400$ & $0.1$ & $0.23,0$ &$0.1$ & $0.1$  & $(0,0.00144)$   \\
        \hline 
        6.(a) & $0.01$ & $(80-11549)$ & $(50-600)$ & $0.1$ & $0.22$ &$0.1$ & $0.1$  & $0$   \\
        \hline         
        6.(b) & $0.01$ & $(80-11549)$ & $(50-600)$ & $0.1$ & $0$ &$0.1$ & $0.1$  & $0.0016$   \\
        \hline   
        6.(c) & $0.5$ & $(80-11549)$ & $(50-600)$ & $0.1$ & $0.23$ &$0.1$ & $0.1$  & $0$   \\
        \hline         
        6.(d) & $0.5$ & $(80-11549)$ & $(50-600)$ & $0.1$ & $0$ &$0.1$ & $0.1$  & $0.00144$   \\
        \hline         
        7 & $(0.01,0.5)$ & $8347$ & $512$ & $0.1$ & $0.1$ &$0.1$ & $0.1$  & $0.0001$   \\
        \hline         
	\end{tabular}
 \caption{Parameter values used for respective figures of the paper. Here, radius of colloids are fixed ar $b=1.0$, $L$ is the linear size of the two-dimensional box (system size), $\phi$ area fraction related to number density, $N$ is the total number of particles and $n_p$ is the pinning fraction. The speed $v_s$ (intrinsic) of a free particle is kept fixed at $v_s = 50$. The time step size $dt=0.01$ are kept fixed for all simulations. The remaining parameters are defined after Eq.\eqref{eq:mainLE}. We have kept $\frac{\lambda_0}{4\pi D_c}=1$ in all the simulations. }
 \label{si_table_params}
\end{table*}

 %%%%%%%%%%%%%%%%%%%%%%%%%%%%%%%%%%%%%%%%%%%%%%%%%%% 
  \begin{figure}[b]
     \centering
     \includegraphics[width=0.32\textwidth]{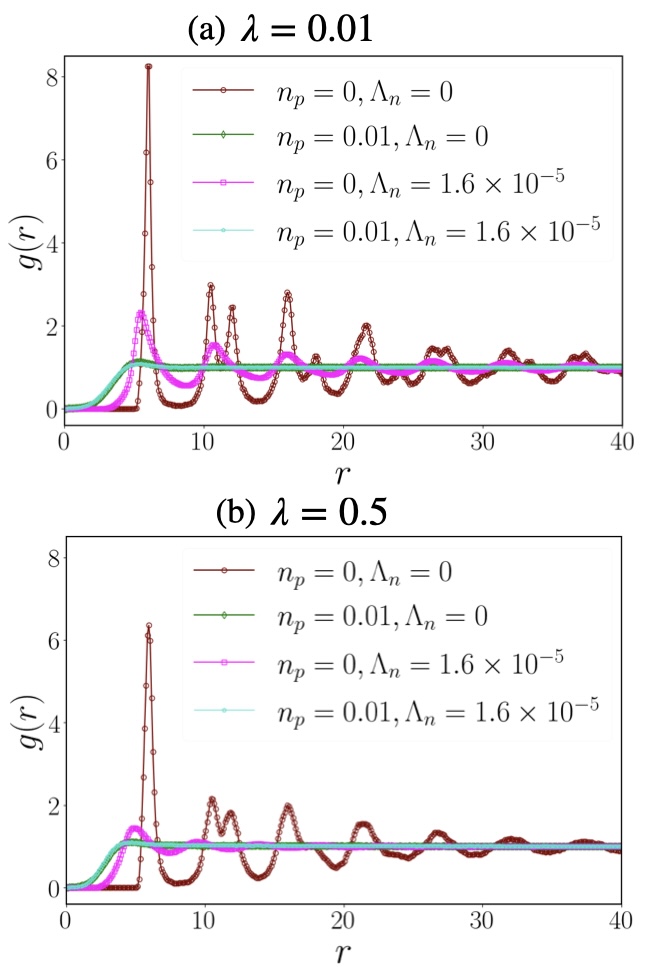}
     \caption{\label{pair_corr}
Pair correlation plot for the co-efficient (a) $\lambda=0.01$ and (b) $\lambda=0.5$. Four cases has been shown for different values of $n_p$ and $\Lambda_n$. System size: $L=200$; $\Lambda_r=\Lambda_t=0.1$; area fraction $\phi=0.1$. }
 \end{figure}
 %%%%%%%%%%%%%%%%%%%%%%%%%%%%%%%%%%%%%%%%%%%%%%%%%%%%%%%%%%%%

\section{Spatial order and pair correlation}
We analyze the radial distribution of particles using the spatial pair correlation function, defined as:
\begin{align}
    g(r) = \frac{1}{N} \sum_{i,j} \delta (r - |\mathbf r_{i} - \mathbf r_{j}|).
\end{align}
It is calculated for the flocking phase at different values of pinning fractions $n_p$ and noises $\Lambda_n$, as shown in Fig.\ref{pair_corr}. The reference parameters used in these simulations are $\phi=0.1$, $\Lambda_t=0.1$ and $\Lambda_r=0.1$. In Fig.\ref{pair_corr}(a) and (b), the pair correlation is calculated for four  cases with different values of disorder and noise, with $\lambda=0.01$ and $\lambda=0.5$, respectively. All other parameter values are kept the same.
In the first case, without disorder ($n_p=0$) and noise ($\Lambda_n=0$), the system forms a well-ordered finite-sized crystalline structure, as evidenced by pronounced peaks in the correlation function (indicated in maroon color). The crystallite phase is stronger in the case of long-range interaction with $\lambda=0.01$. In the second case, noise is $\Lambda_n=0$, a small pinning ($n_p=0.01$) disrupts this order, leading to a transition to a liquid-like phase with reduced peak intensity, which is true for both cases with $\lambda=0.01$ and $0.5$ (indicated in green). In the third case, the is disorder free ($n_p=0$), and a very small amount of noise ($\Lambda_n=1.6 \times 10^{-5}$) is present. The crystallite phase is suppressed comparably with the effective short-range interaction with $\lambda=0.5$. In the fourth case with small disorder and noise, the spatial order is destroyed and fluid-like structure forms, which is similar in both cases as indicated by the turquoise color (merged with the green colored plot). In summary, the spatial order is more robust in the long-range interaction case, especially with the variation of weak noise. \\

 \section{Sample-to-sample fluctuations}
\begin{figure}[t!]
     \centering
     \includegraphics[width=0.32\textwidth]{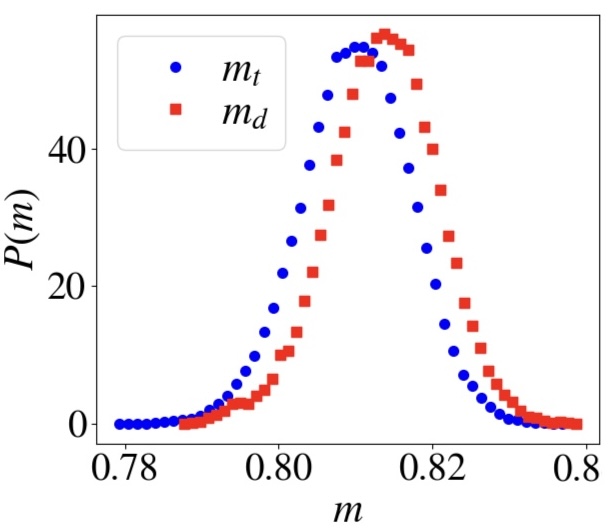}
     \caption{\label{avg}
Order parameter distribution for one realization of disorder over time ($m_t$) and for many realization of disorder ($m_d$) for system of size $L=400$ and area fraction $\phi=0.1$.}  
 \end{figure}

 In active particle systems with local alignment, the presence of quenched disorder has been shown to induce significant sample-to-sample fluctuations, leading to a breakdown of self-averaging \cite{duan2021breakdown}. To examine whether a similar behavior occurs in the present chemorepulsive active system, we compare the distribution of the order parameter obtained from temporal fluctuations within a single disorder realization, $P(m_t)$, with that obtained by averaging over multiple independent disorder realizations, $P(m_d)$. These distributions are shown in Fig.~\ref{avg}(a) for a system of size $L=400$ in the ordered phase, close to the transition. We observe a small but noticeable difference between the two distributions, suggesting that the system exhibits weak non-self-averaging in the presence of pinning disorder.
 %%%%%%%%%%%%%%%%%%%%%%%%%%%%%%%%%%%%%%%%%%%%%%%%%%%%%%%%%%%%
\section{Statistical averages} \label{app:statAvg}
Initially, all particles are uniformly and randomly distributed in space, with their orientations assigned from a uniform random distribution over [$-\pi,\pi$]. A fraction $n_p$ of the particles is then selected uniformly at random and permanently pinned in position. As the system evolves, these pinned particles continue to interact chemically with both pinned and mobile particles but do not undergo translational motion. 
To ensure that the system reaches a steady state, the first $10^{5}$ time steps are discarded, and the subsequent $2 \times 10^{5}$ time steps are used for time averaging. For systems with $L < 400$, statistical averages are further obtained by averaging over $10$ independent disorder realizations, corresponding to a total of $2 \times 10^{6}$ sampled configurations, unless stated otherwise. For the larger system sizes ($L \ge 400$), only four independent disorder realizations are considered, yielding a total of $8 \times 10^{5}$ sampled configurations.
Owing to the long-range nature of the chemorepulsive interactions, the interaction of every particle with all other particles must be evaluated at each time step. Consequently, the computational cost increases rapidly with system size, limiting the number of independent disorder realizations that can be simulated for the largest systems.

 %%%%%%%%%%%%%%%%%%%%%%%%%%%%%%%%%%%%%%%%%%%%%%%%%%%%%%%%%%%%

\bibliography{reference}

@article{torrenegra2024self,
  title={Self-organization of Janus particles: Impact of hydrodynamic interactions in substrate consumption for structure formation},
  author={Torrenegra-Rico, JD and Arango-Restrepo, A and Rubi, JM},
  journal={The Journal of Chemical Physics},
  volume={161},
  number={22},
  year={2024},
  publisher={AIP Publishing}
}

@article{chepizhko2013optimal,
  title={Optimal noise maximizes collective motion in heterogeneous media},
  author={Chepizhko, Oleksandr and Altmann, Eduardo G and Peruani, Fernando},
  journal={Physical review letters},
  volume={110},
  number={23},
  pages={238101},
  year={2013},
  publisher={APS}
}

@article{nelson1979,
  title = {Dislocation-mediated melting in two dimensions},
  author = {Nelson, David R. and Halperin, B. I.},
  journal = {Phys. Rev. B},
  volume = {19},
  issue = {5},
  pages = {2457--2484},
  numpages = {0},
  year = {1979},
  month = {Mar},
  publisher = {American Physical Society},
  doi = {10.1103/PhysRevB.19.2457},
  url = {https://link.aps.org/doi/10.1103/PhysRevB.19.2457}
}

@article{chepizhko2013diffusion,
  title={Diffusion, Subdiffusion, and Trapping of Active Particles in Heterogeneous Media},
  author={Chepizhko, Oleksandr and Peruani, Fernando},
  journal={Physical Review Letters},
  volume={111},
  number={16},
  pages={160604},
  year={2013},
  publisher={APS}
}

@article{loos2023long,
  title={Long-range order and directional defect propagation in the nonreciprocal XY model with vision cone interactions},
  author={Loos, Sarah AM and Klapp, Sabine HL and Martynec, Thomas},
  journal={Physical review letters},
  volume={130},
  number={19},
  pages={198301},
  year={2023},
  publisher={APS}
}

@article{siebers2023exploiting,
  title={Exploiting compositional disorder in collectives of light-driven circle walkers},
  author={Siebers, Frank and Jayaram, Ashreya and Bl{\"u}mler, Peter and Speck, Thomas},
  journal={Science Advances},
  volume={9},
  number={15},
  pages={eadf5443},
  year={2023},
  publisher={American Association for the Advancement of Science}
}

@article{adhikary2025flocking,
  title={Flocking transition in phoretically interacting active particles with pinning disorder},
  author={Adhikary, Sagarika and Subramaniam, Arvin and Singh, Rajesh},
  journal={New Journal of Physics},
  year={2025}
}

@article{ramaswamy2010mechanics,
  title={The mechanics and statistics of active matter},
  author={Ramaswamy, Sriram},
  journal={Annu. Rev. Condens. Matter Phys.},
  volume={1},
  number={1},
  pages={323--345},
  year={2010},
  publisher={Annual Reviews}
}

@article{duan2021breakdown,
  title={Breakdown of ergodicity and self-averaging in polar flocks with quenched disorder},
  author={Duan, Yu and Mahault, Beno{\^\i}t and Ma, Yu-qiang and Shi, Xia-qing and Chat{\'e}, Hugues},
  journal={Physical Review Letters},
  volume={126},
  number={17},
  pages={178001},
  year={2021},
  publisher={APS}
}

@article{tang2025reentrant,
  title={Reentrant phase behavior in binary topological flocks with nonreciprocal alignment},
  author={Tang, Tian and Duan, Yu and Ma, Yu-qiang},
  journal={Physical Review Research},
  volume={7},
  number={2},
  pages={023008},
  year={2025},
  publisher={APS}
}

@article{chate2008collective,
  title={Collective motion of self-propelled particles interacting without cohesion},
  author={Chat{\'e}, Hugues and Ginelli, Francesco and Gr{\'e}goire, Guillaume and Raynaud, Franck},
  journal={Physical Review E—Statistical, Nonlinear, and Soft Matter Physics},
  volume={77},
  number={4},
  pages={046113},
  year={2008},
  publisher={APS}
}

@article{caprini2023flocking,
  title={Flocking without alignment interactions in attractive active Brownian particles},
  author={Caprini, Lorenzo and L{\"o}wen, Hartmut},
  journal={Physical Review Letters},
  volume={130},
  number={14},
  pages={148202},
  year={2023},
  publisher={APS}
}

@article{mokhtari2017collective,
  title={Collective rotations of active particles interacting with obstacles},
  author={Mokhtari, Zahra and Aspelmeier, Timo and Zippelius, Annette},
  journal={Europhysics Letters},
  volume={120},
  number={1},
  pages={14001},
  year={2017},
  publisher={IOP Publishing}
}

@article{adhikary2021effect,
  title={Effect of trapping perturbation on the collective dynamics of self-propelled particles},
  author={Adhikary, Sagarika and Santra, SB},
  journal={Europhysics Letters},
  volume={135},
  number={4},
  pages={48003},
  year={2021},
  publisher={IOP Publishing}
}

@article{subramaniam2025minimal,
  title={A minimal mechanism for flocking in phoretically interacting active particles},
  author={Subramaniam, Arvin Gopal and Adhikary, Sagarika and Singh, Rajesh},
  journal={Soft Matter},
  volume={21},
  number={47},
  pages={9058--9069},
  year={2025},
  publisher={Royal Society of Chemistry}
}

@article{kreienkamp2024dynamical,
  title={Dynamical structures in phase-separating nonreciprocal polar active mixtures},
  author={Kreienkamp, Kim L and Klapp, Sabine HL},
  journal={Physical Review E},
  volume={110},
  number={6},
  pages={064135},
  year={2024},
  publisher={APS}
}

@article{olsen2021active,
  title={Active Brownian particles moving through disordered landscapes},
  author={Olsen, Kristian S and Angheluta, Luiza and Flekk{\o}y, Eirik G},
  journal={Soft Matter},
  volume={17},
  number={8},
  pages={2151--2157},
  year={2021},
  publisher={Royal Society of Chemistry}
}

@article{sandor2017dynamic,
  title={Dynamic phases of active matter systems with quenched disorder},
  author={S{\'a}ndor, Cs and Libal, Andras and Reichhardt, Charles and Olson Reichhardt, Cynthia Jane},
  journal={Physical Review E},
  volume={95},
  number={3},
  pages={032606},
  year={2017},
  publisher={APS}
}

@article{peruani2018cold,
  title={Cold active motion: how time-independent disorder affects the motion of self-propelled agents},
  author={Peruani, Fernando and Aranson, Igor S},
  journal={Physical review letters},
  volume={120},
  number={23},
  pages={238101},
  year={2018},
  publisher={APS}
}

@article{saha2025effervescence,
  title={Effervescence in a binary mixture with nonlinear non-reciprocal interactions},
  author={Saha, Suropriya and Golestanian, Ramin},
  journal={Nature Communications},
  volume={16},
  number={1},
  pages={7310},
  year={2025},
  publisher={Nature Publishing Group UK London}
}

@article{rouzaire2025activity,
  title={Activity leads to topological phase transition in 2D populations of heterogeneous oscillators},
  author={Rouzaire, Ylann and Rahmani, Parisa and Pagonabarraga, Ignacio and Peruani, Fernando and Levis, Demian},
  journal={Physical Review Letters},
  volume={134},
  number={18},
  pages={188301},
  year={2025},
  publisher={APS}
}

@article{baconnier2025self,
  title={Self-aligning polar active matter},
  author={Baconnier, Paul and Dauchot, Olivier and D{\'e}mery, Vincent and D{\"u}ring, Gustavo and Henkes, Silke and Huepe, Cristi{\'a}n and Shee, Amir},
  journal={Reviews of Modern Physics},
  volume={97},
  number={1},
  pages={015007},
  year={2025},
  publisher={APS}
}

@article{bechinger2016active,
  title={Active particles in complex and crowded environments},
  author={Bechinger, Clemens and Di Leonardo, Roberto and L{\"o}wen, Hartmut and Reichhardt, Charles and Volpe, Giorgio and Volpe, Giovanni},
  journal={Reviews of modern physics},
  volume={88},
  number={4},
  pages={045006},
  year={2016},
  publisher={APS}
}

@article{liebchen2019interactions,
  title={Which interactions dominate in active colloids?},
  author={Liebchen, Benno and L{\"o}wen, Hartmut},
  journal={The Journal of chemical physics},
  volume={150},
  number={6},
  year={2019},
  publisher={AIP Publishing}
}

@article{kumar2023emergent,
  title={Emergent dynamics due to chemo-hydrodynamic self-interactions in active polymers},
  author={Kumar, Manoj and Murali, Aniruddh and Subramaniam, Arvin Gopal and Singh, Rajesh and Thutupalli, Shashi},
  journal={Nature Commun.},
  year={2024},
  volume={15},
  number={1},
  pages={4903},
doi={10.1038/s41467-024-49155-7}
}

@book{chaikin1995principles,
  title={Principles of condensed matter physics},
  author={Chaikin, Paul M and Lubensky, Tom C and Witten, Thomas A},
  volume={10},
  year={1995},
  publisher={Cambridge university press Cambridge}
}

@article{vrugt2025exactly,
  title={What exactly is' active matter'?},
  author={Vrugt, Michael te and Liebchen, Benno and Cates, Michael E},
  journal={arXiv preprint arXiv:2507.21621},
  year={2025}
}

@article{hokmabad2022chemotactic,
  title={Chemotactic self-caging in active emulsions},
  author={Hokmabad, Babak Vajdi and Agudo-Canalejo, Jaime and Saha, Suropriya and Golestanian, Ramin and Maass, Corinna C},
  journal={Proc. Natl. Acad. Sci.},
  volume={119},
  number={24},
  pages={e2122269119},
  year={2022},
  publisher={National Acad Sciences}
}

@article{thutupalli2018flow,
  title={Flow-induced phase separation of active particles is controlled by boundary conditions},
  author={Thutupalli, Shashi and Geyer, Delphine and Singh, Rajesh and Adhikari, Ronojoy and Stone, Howard A},
  journal={Proc. Natl. Acad. Sci.},
  volume={115},
  number={21},
  pages={5403--5408},
  year={2018},
  publisher={National Acad Sciences}
}

@article{soto2015self,
  title={Self-assembly of active colloidal molecules with dynamic function},
  author={Soto, Rodrigo and Golestanian, Ramin},
  journal={Phys. Rev. E},
  volume={91},
  number={5},
  pages={052304},
  year={2015},
  publisher={APS}
}

@article{saha2014clusters,
  title={Clusters, asters, and collective oscillations in chemotactic colloids},
  author={Saha, Suropriya and Golestanian, Ramin and Ramaswamy, Sriram},
  journal={Phys. Rev. E},
  volume={89},
  number={6},
  pages={062316},
  year={2014},
  publisher={APS}
}

@article{singh2016universal,
  title={Universal hydrodynamic mechanisms for crystallization in active colloidal suspensions},
  author={Singh, Rajesh and Adhikari, R},
  journal={Phys. Rev. Lett.},
  volume={117},
  number={22},
  pages={228002},
  year={2016},
  publisher={APS}
}

@misc{siText,
  title={See the Supplemental Material at this {URL}: [to be inserted].}
}

@article{ramaswamy2003active,
  title={Active nematics on a substrate: Giant number fluctuations and long-time tails},
  author={Ramaswamy, Sriram and Simha, R Aditi and Toner, John},
  journal={EPL (Europhysics Letters)},
  volume={62},
  number={2},
  pages={196--202},
  year={2003}
}

@article{ginelli2016physics,
  title={The physics of the Vicsek model},
  author={Ginelli, Francesco},
  journal={The European Physical Journal Special Topics},
  volume={225},
  number={11},
  pages={2099--2117},
  year={2016},
  publisher={Springer}
}

@article{narayan2007long,
  title={Long-lived giant number fluctuations in a swarming granular nematic},
  author={Narayan, Vijay and Ramaswamy, Sriram and Menon, Narayanan},
  journal={Science},
  volume={317},
  number={5834},
  pages={105--108},
  year={2007},
  publisher={American Association for the Advancement of Science}
}

@article{chate2006simple,
  title={Simple model for active nematics: quasi-long-range order and giant fluctuations},
  author={Chat{\'e}, Hugues and Ginelli, Francesco and Montagne, Ra{\'u}l},
  journal={Physical review letters},
  volume={96},
  number={18},
  pages={180602},
  year={2006},
  publisher={APS}
}

@article{vicsek1995novel,
  title={Novel type of phase transition in a system of self-driven particles},
  author={Vicsek, Tam{\'a}s and Czir{\'o}k, Andr{\'a}s and Ben-Jacob, Eshel and Cohen, Inon and Shochet, Ofer},
  journal={Physical review letters},
  volume={75},
  number={6},
  pages={1226},
  year={1995},
  publisher={APS}
}

@article{pohlStarkPRL2014,
  title = {Dynamic Clustering and Chemotactic Collapse of Self-Phoretic Active Particles},
  author = {Pohl, Oliver and Stark, Holger},
  journal = {Phys. Rev. Lett.},
  volume = {112},
  issue = {23},
  pages = {238303},
  numpages = {5},
  year = {2014},
  month = {Jun},
  publisher = {American Physical Society},
  doi = {10.1103/PhysRevLett.112.238303},
  url = {https://link.aps.org/doi/10.1103/PhysRevLett.112.238303}
}

@article{jhajhria2025kinetics,
  title={Kinetics of phase transition in nonreciprocal mixtures of passive and chemophoretically active particles},
  author={Jhajhria, Manisha and Das, Subir K and Thakur, Snigdha},
  journal={The Journal of Chemical Physics},
  volume={162},
  number={19},
  year={2025},
  publisher={AIP Publishing}
}

@article{das2024flocking,
  title={Flocking by turning away},
  author={Das, Suchismita and Ciarchi, Matteo and Zhou, Ziqi and Yan, Jing and Zhang, Jie and Alert, Ricard},
  journal={Physical Review X},
  volume={14},
  number={3},
  pages={031008},
  year={2024},
  publisher={APS}
}

@article{adhikary2022pattern,
  title={Pattern formation and phase transition in the collective dynamics of a binary mixture of polar self-propelled particles},
  author={Adhikary, Sagarika and Santra, SB},
  journal={Physical Review E},
  volume={105},
  number={6},
  pages={064612},
  year={2022},
  publisher={APS}
}

@article{chen2024emergent,
  title={Emergent chirality and hyperuniformity in an active mixture with nonreciprocal interactions},
  author={Chen, Jianchao and Lei, Xiaokang and Xiang, Yalun and Duan, Mengyuan and Peng, Xingguang and Zhang, HP},
  journal={Physical Review Letters},
  volume={132},
  number={11},
  pages={118301},
  year={2024},
  publisher={APS}
}

@article{martin2018collective,
  title={Collective motion of active Brownian particles with polar alignment},
  author={Mart{\'\i}n-G{\'o}mez, Aitor and Levis, Demian and D{\'\i}az-Guilera, Albert and Pagonabarraga, Ignacio},
  journal={Soft matter},
  volume={14},
  number={14},
  pages={2610--2618},
  year={2018},
  publisher={Royal Society of Chemistry}
}

@article{sese2018velocity,
  title={Velocity alignment promotes motility-induced phase separation},
  author={Sese-Sansa, Elena and Pagonabarraga, Ignacio and Levis, Demian},
  journal={Europhysics Letters},
  volume={124},
  number={3},
  pages={30004},
  year={2018},
  publisher={IOP Publishing}
}

@article{martin2021statistical,
  title={Statistical mechanics of active Ornstein-Uhlenbeck particles},
  author={Martin, David and O'Byrne, J{\'e}r{\'e}my and Cates, Michael E and Fodor, {\'E}tienne and Nardini, Cesare and Tailleur, Julien and Van Wijland, Fr{\'e}d{\'e}ric},
  journal={Physical Review E},
  volume={103},
  number={3},
  pages={032607},
  year={2021},
  publisher={APS}
}

@article{caprini2022dynamics,
  title={Dynamics of active particles with space-dependent swim velocity},
  author={Caprini, Lorenzo and Marini Bettolo Marconi, Umberto and Wittmann, Ren{\'e} and L{\"o}wen, Hartmut},
  journal={Soft Matter},
  volume={18},
  number={7},
  pages={1412--1422},
  year={2022},
  publisher={The Royal Society of Chemistry}
}

@article{subramaniam2024rigid,
  title={Rigid flocks, undulatory gaits, and chiral foldamers in a chemically active polymer},
  author={Subramaniam, Arvin Gopal and Kumar, Manoj and Thutupalli, Shashi and Singh, Rajesh},
  journal={New Journal of Physics},
  volume={26},
  number={8},
  pages={083009},
  year={2024},
  publisher={IOP Publishing}
}

@article{singh2019competing,
  title={Competing chemical and hydrodynamic interactions in autophoretic colloidal suspensions},
  author={Singh, Rajesh and Adhikari, R and Cates, ME},
  journal={The Journal of chemical physics},
  volume={151},
  number={4},
  year={2019},
  publisher={AIP Publishing}
}

@book{toner2024physics,
  title={The Physics of Flocking: Birth, Death, and Flight in Active Matter},
  author={Toner, John},
  year={2024},
  publisher={Cambridge University Press}
}

@article{marchetti2013hydrodynamics,
  title={Hydrodynamics of soft active matter},
  author={Marchetti, M Cristina and Joanny, Jean-Fran{\c{c}}ois and Ramaswamy, Sriram and Liverpool, Tanniemola B and Prost, Jacques and Rao, Madan and Simha, R Aditi},
  journal={Reviews of modern physics},
  volume={85},
  number={3},
  pages={1143--1189},
  year={2013},
  publisher={APS}
}

@article{ruckner2007chemically,
  title={Chemically powered nanodimers},
  author={R{\"u}ckner, Gunnar and Kapral, Raymond},
  journal={Physical review letters},
  volume={98},
  number={15},
  pages={150603},
  year={2007},
  publisher={APS}
}

@article{matas2014hydrodynamic,
    author={Matas-Navarro, R. and Golestanian, R. and Liverpool, T. B. and Fielding, S. M.},
    journal={Phys. Rev. E},
    volume={90},
    number={3},
    pages={032304},
    year={2014},
    doi={10.1103/PhysRevE.90.032304},
    publisher={APS}
}

@article{gidituri2026microswimmers,
  title={Microswimmers create bicontinuous emulsions in binary fluids},
  author={Gidituri, Harinadha and Samatas, Sotiris and Lintuvuori, Juho S},
  journal={arXiv preprint arXiv:2606.07292},
  year={2026}
}

@article{zhou2026hydrodynamic,
  title={Hydrodynamic interactions destroy motility-induced phase separation in active suspensions},
  author={Zhou, Tingtao and Brady, John F},
  journal={Physical Review Letters},
  volume={136},
  number={8},
  pages={088301},
  year={2026},
  publisher={APS}
}

@article{gompper20252025,
  title={The 2025 motile active matter roadmap},
  author={Gompper, Gerhard and Stone, Howard A and Kurzthaler, Christina and Saintillan, David and Peruani, Fernado and Fedosov, Dmitry A and Auth, Thorsten and Cottin-Bizonne, Cecile and Ybert, Christophe and Cl{\'e}ment, Eric and others},
  journal={Journal of Physics: Condensed Matter},
  volume={37},
  number={14},
  pages={143501},
  year={2025},
  publisher={IOP Publishing}
}

@article{binder1987theory,
  title={Theory of first-order phase transitions},
  author={Binder, Kurt},
  journal={Reports on progress in physics},
  volume={50},
  number={7},
  pages={783},
  year={1987},
  publisher={IOP Publishing}
}

@book{christensen2005complexity,
  title={Complexity and criticality},
  author={Christensen, Kim and Moloney, Nicholas R},
  volume={1},
  year={2005},
  publisher={Imperial College Press}
}

@incollection{janke2003first,
  title={First-order phase transitions},
  author={Janke, W},
  booktitle={Computer Simulations of Surfaces and Interfaces},
  pages={111--135},
  year={2003},
  publisher={Springer}
}

@article{codina2022small,
  title={Small obstacle in a large polar flock},
  author={Codina, Joan and Mahault, Beno{\^\i}t and Chat{\'e}, Hugues and Dobnikar, Jure and Pagonabarraga, Ignacio and Shi, Xia-qing},
  journal={Physical Review Letters},
  volume={128},
  number={21},
  pages={218001},
  year={2022},
  publisher={APS}
}

@article{stoop2018clogging,
  title={Clogging and jamming of colloidal monolayers driven across disordered landscapes},
  author={Stoop, Ralph L and Tierno, Pietro},
  journal={Communications Physics},
  volume={1},
  number={1},
  pages={68},
  year={2018},
  publisher={Nature Publishing Group UK London}
}

@article{das2018polar,
  title={Polar flock in the presence of random quenched rotators},
  author={Das, Rakesh and Kumar, Manoranjan and Mishra, Shradha},
  journal={Physical Review E},
  volume={98},
  number={6},
  pages={060602},
  year={2018},
  publisher={APS}
}

@article{vicsek2012collective,
  title={Collective motion},
  author={Vicsek, Tam{\'a}s and Zafeiris, Anna},
  journal={Physics reports},
  volume={517},
  number={3-4},
  pages={71--140},
  year={2012},
  publisher={Elsevier}
}

@article{morin2017distortion,
  title={Distortion and destruction of colloidal flocks in disordered environments},
  author={Morin, Alexandre and Desreumaux, Nicolas and Caussin, Jean-Baptiste and Bartolo, Denis},
  journal={Nature Physics},
  volume={13},
  number={1},
  pages={63--67},
  year={2017},
  publisher={Nature Publishing Group UK London}
}

@article{hiraiwa2020dynamic,
  title={Dynamic self-organization of idealized migrating cells by contact communication},
  author={Hiraiwa, Tetsuya},
  journal={Physical Review Letters},
  volume={125},
  number={26},
  pages={268104},
  year={2020},
  publisher={APS}
}

@article{grossmann2013self,
  title={Self-propelled particles with selective attraction--repulsion interaction: from microscopic dynamics to coarse-grained theories},
  author={Grossmann, Robert and Schimansky-Geier, Lutz and Romanczuk, Pawel},
  journal={New Journal of Physics},
  volume={15},
  number={8},
  pages={085014},
  year={2013},
  publisher={IOP Publishing}
}

@article{martin2025transition,
  title={Transition to collective motion in nonreciprocal active matter: Coarse graining agent-based models into fluctuating hydrodynamics},
  author={Martin, David and Seara, Daniel and Avni, Yael and Fruchart, Michel and Vitelli, Vincenzo},
  journal={Physical Review X},
  volume={15},
  number={4},
  pages={041015},
  year={2025},
  publisher={APS}
}

@article{solon2015pattern,
  title={Pattern formation in flocking models: A hydrodynamic description},
  author={Solon, Alexandre P and Caussin, Jean-Baptiste and Bartolo, Denis and Chat{\'e}, Hugues and Tailleur, Julien},
  journal={Physical Review E},
  volume={92},
  number={6},
  pages={062111},
  year={2015},
  publisher={APS}
}

@article{gupta2022active,
  title={Active nonreciprocal attraction between motile particles in an elastic medium},
  author={Gupta, Rahul Kumar and Kant, Raushan and Soni, Harsh and Sood, AK and Ramaswamy, Sriram},
  journal={Physical Review E},
  volume={105},
  number={6},
  pages={064602},
  year={2022},
  publisher={APS}
}

@article{shaebani2020computational,
  title={Computational models for active matter},
  author={Shaebani, M Reza and Wysocki, Adam and Winkler, Roland G and Gompper, Gerhard and Rieger, Heiko},
  journal={Nature Reviews Physics},
  volume={2},
  number={4},
  pages={181--199},
  year={2020},
  publisher={Nature Publishing Group UK London}
}

@article{barberis2019phase,
  title={Phase separation and emergence of collective motion in a one-dimensional system of active particles},
  author={Barberis, Lucas and Peruani, Fernando},
  journal={The Journal of chemical physics},
  volume={150},
  number={14},
  year={2019},
  publisher={AIP Publishing}
}

\end{document}